\documentclass[journal=jctcce,manuscript=article]{achemso}

\usepackage{graphicx}
\usepackage{float}
\usepackage{color}
\usepackage{multirow}

\title{Electronic structure of water from Koopmans-compliant functionals}

\author{James Moraes de Almeida}
\affiliation{Universidade Federal do ABC, Centro de Ciências Naturais e Humanas, Santo André, São Paulo, Brazil}
\alsoaffiliation{Theory and Simulation of Materials (THEOS) and National Centre for Computational Design and Discovery of Novel Materials (MARVEL), Ecole Polytechnique F\'ed\'erale de Lausanne (EPFL), CH-1015 Lausanne, Switzerland}
\email{james.almeida@ufabc.edu.br}
\author{Ngoc Linh Nguyen}
\affiliation{Theory and Simulation of Materials (THEOS) and National Centre for Computational Design and Discovery of Novel Materials (MARVEL), Ecole Polytechnique F\'ed\'erale de Lausanne (EPFL), CH-1015 Lausanne, Switzerland}
\author{Nicola Colonna}
\affiliation{Laboratory for Neutron Scattering and Imaging, Paul Scherrer Institute, CH-5232 Villigen-PSI, Switzerland}
\affiliation{Theory and Simulation of Materials (THEOS) and National Centre for Computational Design and Discovery of Novel Materials (MARVEL), Ecole Polytechnique F\'ed\'erale de Lausanne (EPFL), CH-1015 Lausanne, Switzerland}
\author{Wei Chen}
\affiliation{Institute of Condensed Matter and Nanoscience, Université catholique de Louvain, B-1348 Louvain-la-Neuve, Belgium}
\author{Caetano Rodrigues Miranda}
\affiliation{Instituto de F\'isica, Universidade de S\~ao Paulo, S\~ao Paulo, SP, Brazil}
\author{Alfredo Pasquarello}
\affiliation{Chaire de Simulation \`a l'Echelle Atomique (CSEA), Ecole Polytechnique F\'ed\'erale de Lausanne (EPFL), CH-1015 Lausanne, Switzerland}
\author{Nicola Marzari}
\affiliation{Theory and Simulation of Materials (THEOS) and National Centre for Computational Design and Discovery of Novel Materials (MARVEL), Ecole Polytechnique F\'ed\'erale de Lausanne (EPFL), CH-1015 Lausanne, Switzerland}

\begin{document}
\newpage
\begin{abstract}
Obtaining a precise theoretical description of the spectral properties of liquid water poses challenges for both molecular dynamics (MD) and electronic structure methods.  
The lower computational cost of the Koopmans-compliant functionals with respect to Green's function methods allows the simulations of many MD trajectories, with a description close to the state-of-art quasi-particle self-consistent GW plus vertex corrections method (QSGW+f$_{xc}$). Thus, we explore water spectral properties when different MD approaches are used, ranging from classical MD to first-principles MD, and including nuclear quantum effects.  
We have observed that the different MD approaches lead to up to 1 eV change in the average band gap, thus, we focused on the band gap dependence with the geometrical properties of the system to explain such spread. 
We have evaluated the changes in the band gap due to variations in the intramolecular O-H bond distance, and HOH angle, as well as the intermolecular hydrogen bond O$\cdot\cdot\cdot$O distance, and the OHO angles. We have observed that the dominant contribution comes from the O-H bond length; the O$\cdot\cdot\cdot$O distance plays a secondary role, and the other geometrical properties do not significantly influence the gap. Furthermore, we analyze the electronic density of states (DOS), where the KIPZ functional shows a good agreement with the DOS obtained with state-of-art approaches employing quasi-particle self-consistent GW plus vertex corrections. The O-H bond length also significantly influences the DOS. When nuclear quantum effects are considered, a broadening of the peaks driven by the broader distribution of the O-H bond lengths is observed, leading to a closer agreement with the experimental photoemission spectra. 
\end{abstract}
\newpage
\maketitle

\section*{Introduction}

Water is a fundamental component of life and present in abundance on our planet. Despite its apparently simple molecular structure, in its condensed form, water shows a very complex behavior
that still challenges researchers across multidisciplinary areas~\cite{schwegler2004towards, morrone2008nuclear, kuo2004liquid, cicero2008water, sprik1996ab, lippert1999j, fois1994properties, PhysRevLett.96.215502, molinero2009water, moore2011structural}. Obtaining the precise photoelectronic properties of liquid water is an active research area\cite{PhysRevLett.117.186401,PhysRevB.89.060202, C4CP04202F, doi:10.1063/1.1979487, rozsa2018ab, gaiduk2018first, pham2017modelling}. In particular, it is an essential issue for electrocatalytic\cite{doi:10.1021/acs.accounts.6b00001,ADVS:ADVS255, rossmeisl2005electrolysis} and photocatalytic water splitting\cite{photocat, mckone2014will, pham2017modelling}, which have a direct impact on hydrogen generation from solar energy\cite{Radecka200846}. Improved knowledge on the photoelectronic properties of water involves some unique complexities of this deceivingly simple system. Thus, there are many challenges both for experimental and theoretical methods.

On the experimental side, the water valence band maximum (VBM) can be determined from photoemission spectroscopy. A value of 9.9 eV below the vacuum level is obtained by linearly extrapolating the slope of the 1b$_1$ level near the photoionization threshold\cite{doi:10.1021/jp030263q}. One can obtain the VBM value with other techniques, yielding  different values\cite{PhysRevB.89.060202, delahay1981photoelectron}, but all clustered around 10 eV\cite{gaiduk2018electron}. The measurement of the conduction band minimum (CBM) is more involved. The main challenge is that an excess electron in the water does not occupy the CBM level; it will go to deeper trapped states\cite{excesselec1, abel2012nature, herbert2017hydrated, marsalek2012structure, uhlig2012unraveling, boero2003first, kumar2015simple, ambrosio2017electronic}. For example, without considering the trapped states, the CBM level obtained by photoelectrochemical methods is at -1.2 eV from the vacuum level\cite{GRAND197973,doi:10.1021/j100447a039}, whereas when considering the trapped states it becomes -0.74 eV\cite{BERNAS1997151}. There is also a recent debate on the energy level this excess electron occupies\cite{doi:10.1021/jacs.6b06715,Luckhause1603224}. Hence, one can model the CBM level from the experimental data obtained for the excess electron with thermodynamical constraints, as performed by Ambrosio \textit{et al.} \cite{doi:10.1021/acs.jpclett.7b00699}, obtaining a CBM value of 0.97 eV below the vacuum level, with many other values found in the literature\cite{coons2016hydrated, bernas1997electronic, coe1997using, grand1979photoionization, coe2001fundamental, donald2010electron, stahler2015real, king2017trapped, pham2014probing, fang2015accurate, ziaei2017dynamical}. Taking into account the various values from the literature, the typical value for the band gap is estimated to be 8.7 $\pm$ 0.6 eV\cite{PhysRevLett.117.186401,PhysRevB.89.060202}.

From a theoretical point of view, the modeling of water brings several challenges. The first one is the need to perform careful molecular dynamics (MD) sampling using classical or \textit{first-principles} MD\cite{PhysRevB.89.060202} eventually including nuclear quantum effects~\cite{PhysRevLett.117.186401}. Furthermore, depending on the electronic structure method employed (ranging from density functional theory to self-consistent GW calculations with efficient vertex corrections), one can obtain for the same trajectory different average band gaps\cite{PhysRevLett.117.186401}. Thus, it is not simple to clearly understand what drives the band gap changes in different trajectories. 
In this work, we study the electronic structure of liquid water using Koopmans-compliant (KC) spectral functionals\cite{PhysRevB.82.115121,PhysRevB.90.075135,PhysRevB.89.195134}; which have already shown to be a valid and reliable alternative to more complex techniques for the calculation of photoemission properties of different molecular or extended systems~\cite{PhysRevB.89.195134,doi:10.1021/acs.jctc.6b00145,PhysRevLett.114.166405,colonna_screening_2018,nguyen_koopmans-compliant_2018,colonna_koopmans-compliant_2019,elliott_koopmans_2019}. We take advantage of the lower computational cost of the KC functionals, when compared to many-body methods, to calculate the electronic properties of liquid water over different MD trajectories, ranging from classical MD to first-principles MD trajectories, eventually including nuclear quantum effects. We analyze the effect of geometrical properties on the band gap elucidating the relation between electronic structure and average O-H bond length, OHO angle, O$\cdot\cdot\cdot$O hydrogen bond distances, and OHO hydrogen bond angle. 
Additionally, we also compare the density of states obtained from different trajectories with the experimental photoemission spectra, correlating different features with the average geometrical properties of water.

\section*{Methodology}

We have used trajectories for liquid water at 300 K and 1 g/cm$^3$ obtained with either classical or \textit{first-principles} MD. The classical trajectories were obtained using the LAMMPS package\cite{lammpsref}, with commonly used SPCE/FH \cite{:/content/aip/journal/jcp/130/17/10.1063/1.3124184} and TIP4P\cite{doi:10.1063/1.2121687} water potentials. In both cases we used a simple-cubic supercell with 64 water molecules. The equilibration run (NVT, 300 K, 1 g/cm$^3$) was 5 ns long, followed by a production run of 10 ns. For the electronic structure calculations, we took 20 samples from the production run, 0.5 ns apart.

In this work, we use the first-principles MD trajectories generated in Ref.~\citenum{PhysRevLett.117.186401}. These simulations were performed with the PWSCF code of the Quantum ESPRESSO distribution\cite{0953-8984-21-39-395502,giannozzi_advanced_2017} and the i-PI wrapper\cite{Ceriotti20141019} for the nuclear degrees of freedom. The exchange-correlation functional was the revised Vydrov and Van Voorhis (rVV10)\cite{doi:10.1063/1.3521275, PhysRevB.87.041108}, with a short-ranged parameter tuned to obtain a 1 g/cm$^3$ density at 300 K and 1 atm, following Refs. \citenum{PhysRevLett.117.186401} and \citenum{doi:10.1063/1.4938189}. The simulation cell was a simple-cubic with 32 water molecules. SG15 Optimized Norm-Conserving Vanderbilt pseudopotentials\cite{PhysRevB.88.085117} were used to model the electron-nucleus interaction. The kinetic cutoff energy used was 85 Ry. The trajectories with classical nuclei had an equilibration time of 5 ps and a production run of 10 ps. For the present study, we use 20 samples for the electronic structure calculations, each 0.5 ps apart. 

From the final classical nucleus configuration with 32 water molecules, a run including nuclear quantum effects (nqe) is performed, using a generalized Langevin equation thermostat\cite{PhysRevLett.103.030603}. This setup ensures convergence of the path integral mapping onto a classical ring polymer system with only six beads~\cite{quantumbeads,doi:10.1063/1.4772676,doi:10.1063/1.3556661}. After an equilibration time of 2 ps, we let the system evolve for a production run of 10 ps. We finally selected 20 samples for the electronic structure calculations at every 0.5 ps. Since each step of the trajectory is mapped onto six different classical systems (bead), we have, for 20 snapshots, a total of 120 atomic configurations.

We performed the DFT and KC electronic structure calculations using a modified version of the CP code of the Quantum ESPRESSO distribution\cite{0953-8984-21-39-395502, giannozzi_advanced_2017}. The Perdew-Burke-Ernzerhof exchange-correlation functional\cite{PhysRevLett.77.3865} was used for the standard DFT calculations and as the ''base'' density functional for KC calculations. The plane-wave cutoff was 80 Ry.
As a reminder, KC functionals aim at correctly describing charged excitations by enforcing  a generalized piece-wise linear condition of the energy as a function of the occupation $f_i$ of any orbital $\phi_i$ in the system\cite{PhysRevB.23.5048, KC1, KC2, KC3, KC4, doi:10.1021/acs.jctc.6b00145, colonna_screening_2018,PhysRevX.8.021051}. This is achieved augmenting any approximate density functional with simple orbital density dependent corrections $\Pi_i^{KC}[\rho_i]$:

\begin{equation}
E^{KC}[\rho,{\rho_i}] = E^{DFT}[\rho]+\sum_i \alpha_i \Pi_i^{KC}[\rho,\rho_i],
\end{equation}

\begin{equation}
\Pi_i^{KC}[\rho,\rho_i]=-\int_0^{f_i} \langle \phi_i|\hat{H}^{DFT}(s)|\phi_i\rangle ds+f_i\eta_i
\end{equation}

Where $\rho_i(r)=f_i n_i(r) = f_i |\phi_i(r)|^2$ and $\rho=\sum_i\rho_i(r)$, and $\hat{H}^{DFT}(s)$ is the KS Hamiltonian with a fractional occupation $s$ in orbital $\phi_i$. The corrective term $\Pi_i^{KC}$ removes the non-linear behaviour of the approximate density functional and replaces it with a linear one. Depending on the slope of the latter, different flavours of KC functionals can be defined. In the KI functional the slope is chosen as the DFT energy difference between two adjacent integer point and reads:

\begin{equation}
\eta_i^{KI} = E^{DFT}[f_i=1]-E^{DFT}[f_i=0]=\int_0^1\langle \phi_i|\hat{H}^{DFT}(s)|\phi_i\rangle ds,
\end{equation}

thus, the KI functionals alters only the spectra, but preserves the energetics of the underlying density functional. The KIPZ functional adds to KI a Perdew-Zunger (PZ) self-interaction correction (SIC) term, and makes the functional exact for any one-electron system:

\begin{equation}
\eta_i^{KIPZ} = E^{PZ}[f_i=1] - E^{PZ}[f_i=0]=\int_0^1\langle \phi_i|\hat{H}^{PZ}(s)|\phi_i\rangle ds,
\end{equation}

where $\hat{H}^{PZ}(s)=\hat{H}^{DFT}(s)-\hat{\nu}^{DFT}_{H_{XC}}[s|\phi_i^2|]$, with the latter term being the PZ-SIC. Screening and relaxation effects that naturally happen when adding/removing a particle to the system are accounted for by the orbital dependent screening coefficient $\alpha_i$.\cite{colonna_screening_2018, nguyen_koopmans-compliant_2018}

\section*{Results}

\begin{figure}[ht]
\includegraphics[width=1.00\textwidth]{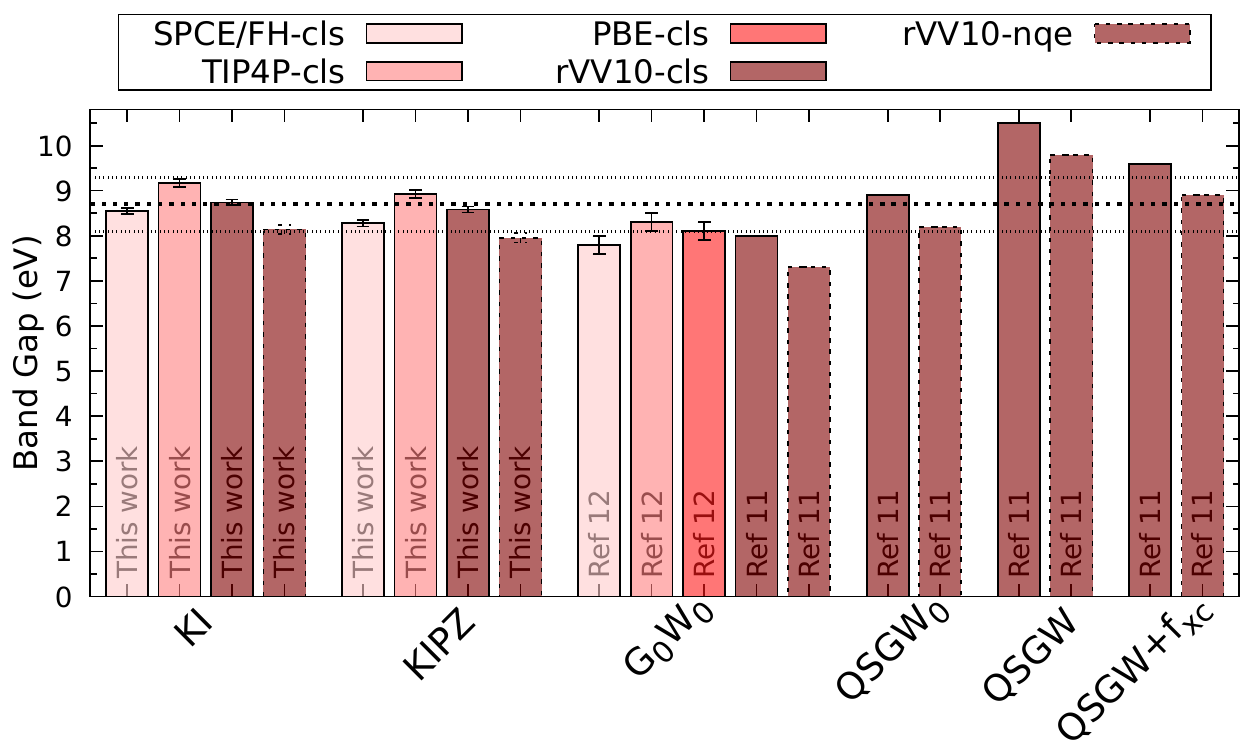}
\caption{\label{fig:geometries} Water band gaps for the five MD approaches and electronic structure methods considered. The inset indicates the classical or first-principles simulations, where in addition ``cls'' stand for classical nuclei and ``nqe'' for nuclear quantum effects. The x-axis label indicates the electronic structure method. The dashed black line is the experimental value, and the black dotted lines are the experimental gap $\pm$ the associated error.}
\end{figure}

The valence band maximum (VBM) and conduction band minimum (CBM) were obtained using a linear extrapolation of the electronic density of states (DOS) near the band edges\cite{doi:10.1063/1.4938189}. This procedure allows to obtain reliable levels even with a water box of only 32 molecules,\cite{doi:10.1063/1.4938189, PhysRevLett.117.186401} and it is similar to the experimental procedure used to  obtain the VBM level from the photoemission (PE) spectra\cite{doi:10.1021/jp030263q}. We show the calculated band gaps in Fig.~\ref{fig:geometries}, comparing also our results with several GW calculations in the literature~\cite{PhysRevLett.117.186401,PhysRevB.89.060202}. We find that the KI (8.14 eV) and KIPZ (7.96 eV) results are very close to the QSGW$_0$ ones when comparing the same trajectories, i.e. first-principles MD trajectories computed with the rVV10 functional either treating the nuclei as classical (rVV10-cls) or with path integral formulation (rVV10-nqe). When we consider the nuclear quantum effects, the KI and KIPZ gaps are decreased by 0.6 eV (on the rVV10 trajectory); these values are close to the 0.7 eV decrease obtained at QSGW+f$_{xc}$ level\cite{PhysRevLett.117.186401}. We have also analyzed the band edge values for the KI functional using the rVV10 trajectories, as can be seen on fig. S1 of Supplementary Material, and the values are in agreement with various GW calculations\cite{gaiduk2018electron,PhysRevLett.117.186401}, being the VBM between the G$_0$W$_0$ and QSGW$_0$, and the CBM closer to the QSGW and QSGW+f$_XC$ results. In addition, we have analyzed the variation of the band edge levels when going from a classical nuclei rVV10 trajectory to the nqe one (Figure S1). For the VBM we have observed an increase of 0.38 eV, whereas for the CBM, a decrease of 0.16 eV. These values compare well with the VBM increase of 0.3 eV and the CBM decrease of 0.2 eV observed in the G$_0$W$_0$ calculations of Gaiduk et al.\cite{gaiduk2018electron}. Our variations are also in agreement with the GW results from Chen et al.\cite{PhysRevLett.117.186401}, as one can see in table S1 of the supplementary material.

\begin{figure}[htbp]
\begin{center}
\includegraphics[width=0.59\textwidth]{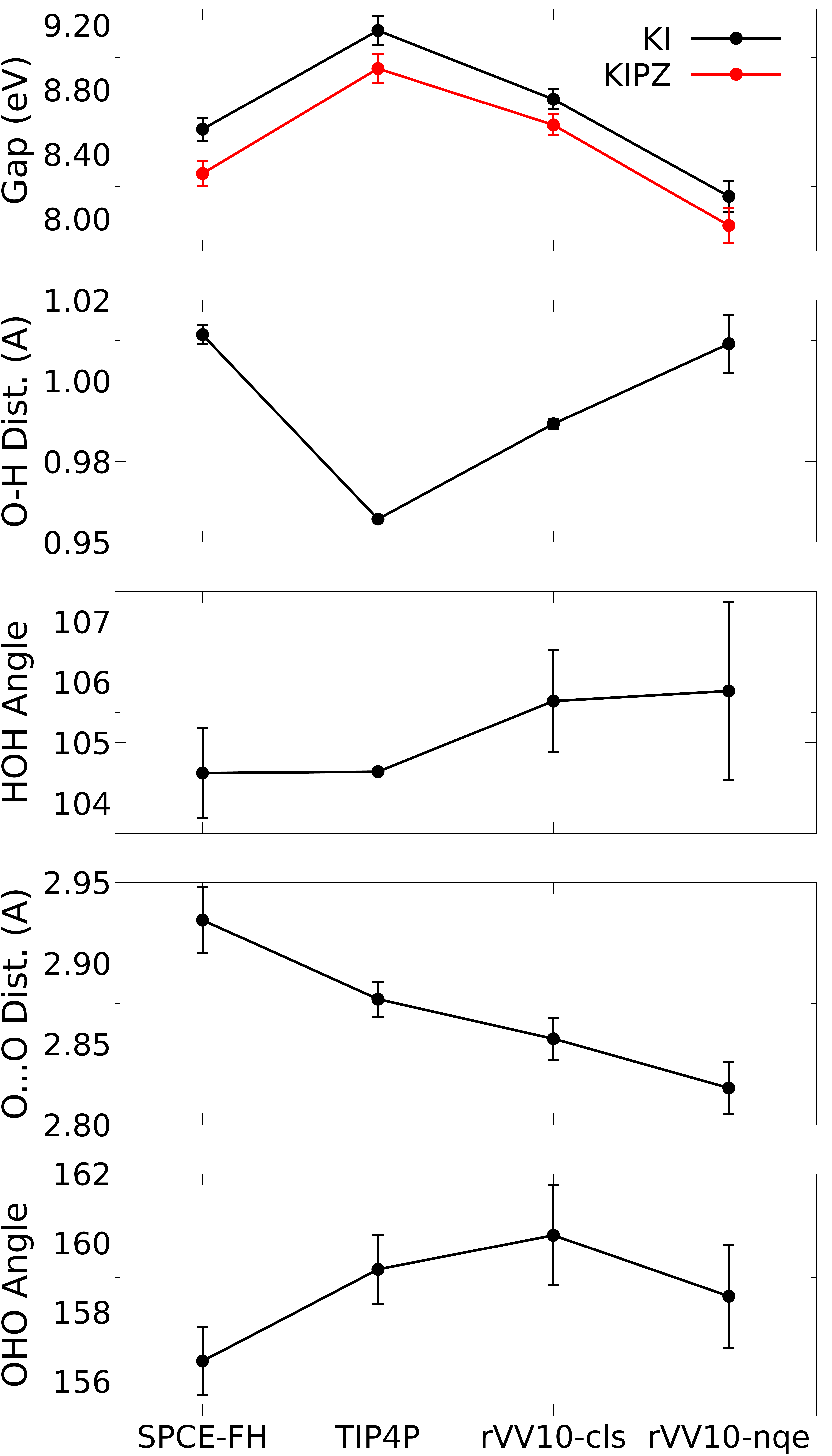}
\caption{\label{fig:variations} Panel a) KI and KIPZ band gaps, b) O-H internal bond lengths, c) HOH internal angles, d) O$\cdot\cdot\cdot$O hydrogen bond distances, and e) OHO hydrogen bond angles for the trajectories considered.}
\end{center}
\end{figure}

When taking into account the band gaps obtained with KI and KIPZ for all the trajectories considered, the difference between the highest and lower band gaps obtained is around 1 eV. To understand these differences, we analyze the geometrical properties of each trajectory, and we plot our results for all considered trajectories and their average geometries in Fig.~\ref{fig:variations}. One can see that the band gap variations are inversely related to the internal O-H bond length (Fig.~\ref{fig:variations} a) and b)). In order to quantify this dependence, we took one snapshot of the rVV10-cls trajectory and then rescaled the internal O-H bond length. The dependence of the gap on the bond length is linear, as can be seen in Fig. S2 in Supplementary Material. This allows for a quantitative estimation of the gap changes due to the O-H bond length variation. 
In addition, the band edge levels variations were analyzed, and shown in Fig. S3. One can see that the KI and KIPZ variations are very similar. By performing a linear fit of the variations, we find that the slopes of the CBM are about 80\% larger than the VBM ones. Thus, the main contributing factor for the gap changes comes from the CBM, although, the contribution from the VBM cannot be neglected.
However, from Fig.~\ref{fig:variations} no direct insight can be obtained for the internal HOH angle, hydrogen bond O$\cdot\cdot\cdot$O distance, and OHO angles.

One cannot isolate the effects of each geometry variation in the bulk liquid water; for example rescaling the O$\cdot\cdot\cdot$O hydrogen bond distance would also alter the density of the system, which also influences the band gap\cite{PhysRevB.89.060202}. Moreover, the hydrogen bond angles cannot be altered collectively: one water molecule has several hydrogen bonds, and rotating the molecule will change all its OHO angles, increasing some and decreasing others. We focus here on the water dimer as a model system to understand how the hydrogen bond geometrical properties influence the HOMO-LUMO gap when varying the HOH internal angle, O$\cdot\cdot\cdot$O hydrogen bond distances, and OHO hydrogen bond angle. The obtained HOMO-LUMO gaps are shown in Fig. S4 (Supplementary Material). We observed that the HOH internal angle and the OHO hydrogen bond angle both have a small influence on the HOMO-LUMO gap of the dimer. However, the O$\cdot\cdot\cdot$O hydrogen bond distance has some contribution to the dimer's HOMO-LUMO gap, with a linear variation. 
Thus, one can focus only on the O-H bond distance and the O$\cdot\cdot\cdot$O hydrogen bond distance to understand how the liquid water band gap varies. As we have shown that both distance variations are linear, we have performed a linear regression on the KIPZ band gap of all calculated snapshots of liquid water, with the following equation:
\begin{equation}
    E_g(d_{O-H},d_{O \cdot\cdot\cdot O})=A*d_{O-H}+B*d_{O\cdot\cdot\cdot O}+C ,
\end{equation}
where $E_g(d_{O-H},d_{O \cdot\cdot\cdot O})$ is the band gap of liquid water, $d_{O-H}$ is the average O-H bond distance, and $d_{O\cdot\cdot\cdot O}$ is the average hydrogen bond distance. A, B, and C are the linear regression coefficients for which we have obtained the following values: A=-14.88 $\pm$ 0.86 eV/\AA, B=1.29  $\pm$ 0.45 eV/\AA, and C=19.69 $\pm$ 1.6 eV. Using these coefficients and the average values of O-H and O$\cdot\cdot\cdot$O from our trajectories, we can obtain an estimation of the band gaps that can be directly compared with the calculated values in order to check the reliability of the fit and of the functional form. As can be seen in Table~\ref{tab:linearfit} the errors associated to the gap obtained from the linear fit are small, with the largest value of 4.0\% for the rVV10-nqe trajectory. Hence, the fit reasonably reproduces the calculated band gaps.

\begin{table*}
\centering
\caption{\label{tab:linearfit} Comparison between the gaps obtained from the linear fit (5th column) and calculated (2nd column) band gaps for all trajectories considered, with the 6th Column showing the relative error. 
}
\begin{tabular}{|c|c|c|c|c|c|} \hline
\multirow{2}{*}{Trajectory} & KIPZ     & O-H Bond     & O$\cdot\cdot\cdot$O & Linear Fit &           \\ 
                            & Gap (eV) & Length (\AA) & Distance (\AA)      & Gap (eV)   &$|$Error$|$\\ \hline
                 SPCE/FH    & 8.28     & 1.014       & 2.927              & 8.37       &   1.1\%  \\ \hline 
                 TIP4P      & 8.93     & 0.957       & 2.878              & 9.15       &   2.5\%  \\ \hline 
                 rVV10-cls  & 8.58     & 0.987       & 2.853              & 8.68       &   1.2\%  \\ \hline 
                 rVV10-nqe  & 7.96     & 1.012       & 2.823              & 8.27       &   4.0\%  \\ \hline 
\end{tabular}
\end{table*}

With A and B obtained from the linear regression, one can see that $d_{O-H}$ has more influence on the band gap then $d_{O\cdot\cdot\cdot O}$, as A is 11.5 times larger than B. As can be seen in table \ref{tab:linearfit}, the difference between the largest and the lowest average $d_{O-H}$ of our trajectories is 0.057 \AA, while for $d_{O\cdot\cdot\cdot O}$ it is 0.104 \AA, thus, the variations on the hydrogen bond distance are larger than on the internal bond length. However, when multiplying $d_{O-H}$ and $d_{O\cdot\cdot\cdot O}$ by A and B, respectively, we observe an absolute gap variation of 0.85 eV from $d_{O-H}$ and 0.13 eV from $d_{O\cdot\cdot\cdot O}$. Thus, the main driving force for changes in the band gap are the internal bond distances, as they contribute to a change in the band gap 6.5 times larger than the hydrogen bond distance. This analysis suggests that the intermolecular properties of the liquid water play a minor role when the water band gap is considered and, as a first approximation, liquid water can be regarded as a collection of molecules as long as its electronic properties are considered. 

\begin{figure}[htbp]
\begin{center}
\includegraphics[width=1\textwidth]{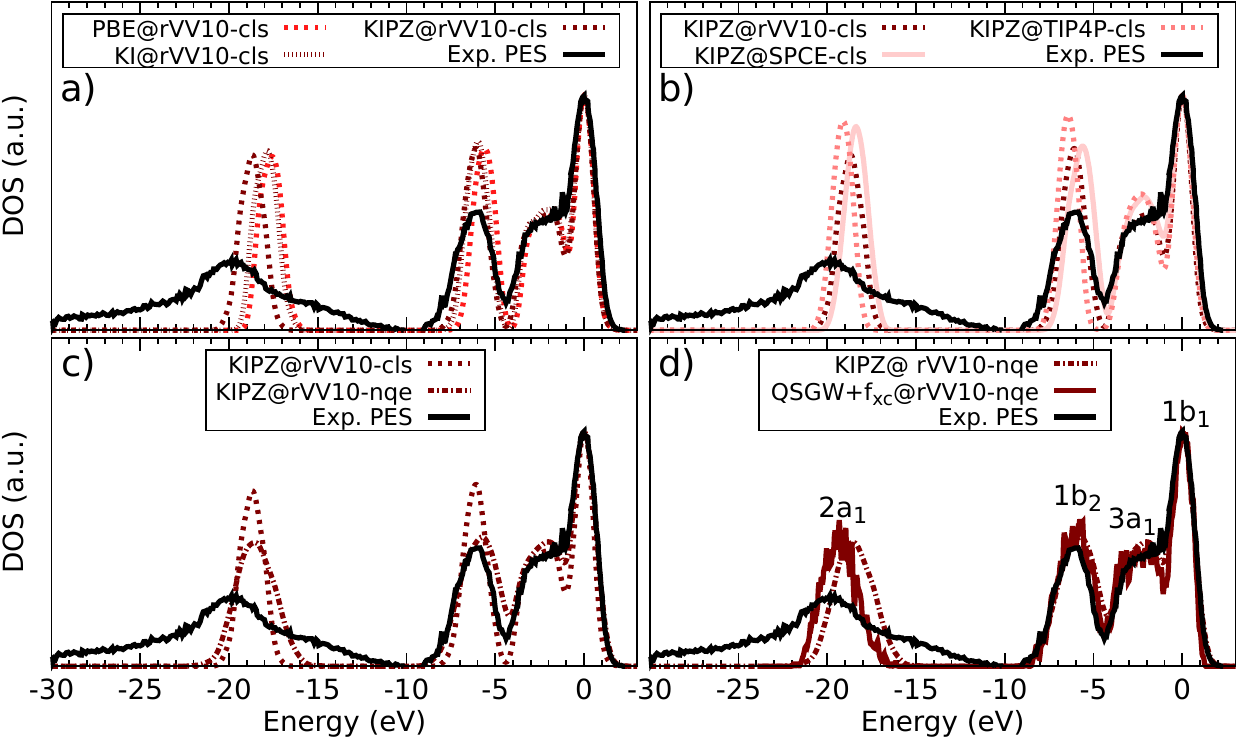}
\caption{\label{fig:dos-multiplot} The density of states (DOS) obtained with the different methodologies are plotted and compared to the experimental photoemission spectra (PES)\cite{winter2004full}. All values are aligned and normalized by the 1b$_1$ peak.}
\end{center}
\end{figure}

Regarding the spectral properties, we show in Fig.~\ref{fig:dos-multiplot} the density of states (DOS) for the different trajectories and methodologies considered. In panel a), the PBE, KI, and KIPZ functionals results for the rVV10-cls trajectory are shown, and compared to the experimental photoemission spectra (PES)\cite{winter2004full}. It can be seen that the KI functional does not alter much the shape of the DOS as compared to PBE (apart, of course, for the opening of the band gap). Instead, KIPZ provides a closer agreement with the PES peak positions, especially for the 2a$_1$ peak. In panel b), the KIPZ results for the classical MD trajectories are compared to those of rVV10-cls. One can see that the 2a$_1$ peak differs significantly. Since the SPCE/FH and TIP4P potentials have an average O-H bond length of 1.01 \AA, and 0.957 \AA, respectively, it indicates that again the O-H distance might be driving the changes in the spectra.
In panel c) of fig.~\ref{fig:dos-multiplot} the KIPZ@rVV10-cls and KIPZ@rVV10-nqe results are compared to the experimental PES to highlight the role of nuclear quantum effects. One can see that the 2a$_1$ peak becomes broader and lower in intensity compared to the one obtained neglecting nqe, whereas the peak position remains instead almost unchanged. This broadening happens because of the larger spread in the O-H bond lengths due to the nqe and it leads to a closer agreement with the experimental PES. Although including nqe leads to a better description of the 2a$_1$ peak shape, the results still show discrepancies with experiments. The main 2a$_1$ peak has two satellite peaks which have been assigned to be due to secondary photoemission processes from other orbitals\cite{doi:10.1021/jp030263q}; which cannot be described within a single-particle theory\cite{doi:10.1021/acs.jctc.6b00630}. The 1b$_2$ peak shows an intensity quite close to the one from the measured PES; however, the peak position is slightly pushed away from the measured position when one moves from the classical to the quantum description of the nuclei. The 3a$_1$ peak shape is in much better agreement with the PES, and also the valley between 1b$_1$ and 3a$_1$ is better described. In panel d) of Fig.~\ref{fig:dos-multiplot} the KIPZ@rVV10-nqe is compared to the DOS obtained with QSGW+f$_{xc}$@rVV10-nqe (from Ref.~\citenum{PhysRevLett.117.186401}) and the experimental PES. The KIPZ results are in good agreement with the QSGW+f$_{xc}$@rVV10-nqe, though the 2a$_1$ peak has a lower energy when using QSGW+f$_{xc}$. Interestingly, the valley between 1b$_1$ and 3a$_1$ is deeper for QSGW+f$_{xc}$. Hence, the inclusion of nqe is crucial for the description of the water photoemission spectra not only because of the longer average O-H bond lengths, but also because of the larger spreads that broaden the DOS peaks and that lead to shallower valleys. 

\begin{figure}[htbp]
\begin{center}
\includegraphics[width=0.99\textwidth]{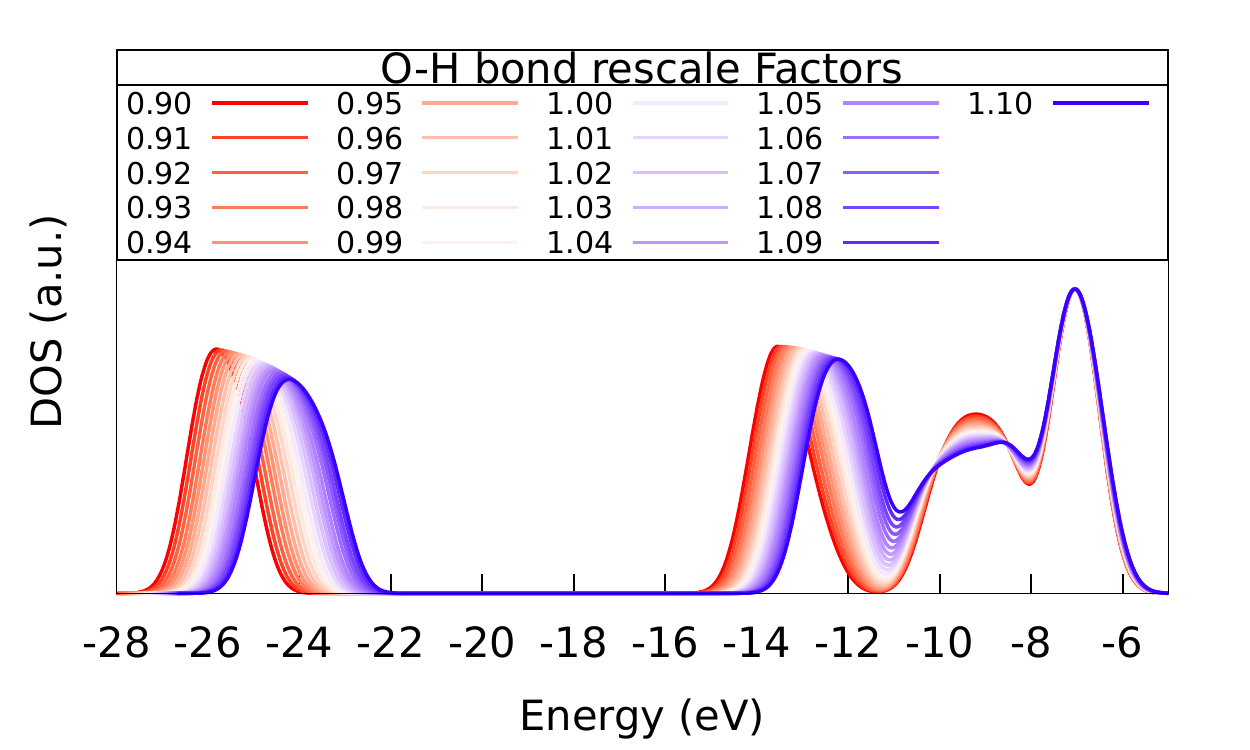}
\caption{\label{fig:dos-rescale} The electronic density of states obtained for a snapshot with the O-H bond distances rescaled from 0.90 to 1.10 times the original bond length.}
\end{center}
\end{figure}

Finally, to explore in more detail the effect of the O-H bond length changes on the DOS, we refer again to the rescaled O-H bond length simulation. The DOS for the rescaled O-H bond length is shown in Fig. \ref{fig:dos-rescale}, which indicates that as the O-H bond length gets smaller the 2a$_1$ and 1b$_2$ peaks move to lower energies. The 3a$_1$ peak becomes instead flatter for longer O-H bond lengths, and the valley between 1b$_1$ and 3a$_1$ becomes shallower. Thus, TIP4P simulations display deeper 2a$_1$ and 1b$_2$ peaks because of the shorter O-H bond lengths, while the opposite is observed in the SPCE/FH simulations.

\section*{Conclusions}

The KI and KIPZ functionals showed consistent performance in obtaining the band gap of liquid water. The results are in good agreement with the experimental value of 8.7 $\pm$ 0.6 eV, with 8.14 eV for KI and 7.96 eV for KIPZ, when using snapshots taken from the rVV10 trajectory including nuclear quantum effects. When comparing with the GW results, KI and KIPZ performances are very similar to those of QSGW$_0$, with snapshots from the same trajectory. The QSGW+f$_{xc}$ yields a higher band gap, 8.9 eV, for the rVV10 trajectory, including nuclear quantum effects.

Band gaps obtained with the same electronic structure methodology, but different MD trajectories can differ by as much as 1 eV. This change is quite striking, and we were able to show that it is mainly related to the geometry of the system, with the O-H bond length being the most dominant factor.
The hydrogen bond lengths played a secondary role, while the internal and hydrogen bond angles barely influenced the band gap of liquid water.

We have compared the DOS of different methods/trajectories with the experimental PES. Apart from a significant improvement of the band gap, the KI functional did not improve much the DOS compared to standard DFT results, whereas the KIPZ provided better results, especially for the 2a$_1$ peak position. The DOS obtained with SPCE/FH and TIP4P indicates that the O-H bond length also plays a major role in the spectral properties, as they showed significant changes in the peaks positions, and they also have very different O-H bond lengths. Also, we showed that when rescaling the O-H bond distance of a fixed snapshot, the DOS can change significantly. This evidence explains the broadening of the DOS peaks observed when the nuclear quantum effects are considered, since they lead to a larger uncertainty in the O-H bond lengths. Overall, the performance of the KIPZ functional is close to the one of QSGW+f$_{xc}$ for the photoemission spectra. In particular, KIPZ provides a very good description of the valley between the 1b$_1$ and 3a$_1$ peaks but an underestimation of the binding energy of the 2a$_1$ peak by 0.74 eV compared to QSGW+f$_{xc}$. 

In conclusion, we are able to tackle the photoelectronic properties variations in liquid water, correlating the influence of the geometrical properties on both the band gap and the density of states. Our data indicate that the spectral properties of water are mainly influenced by the intrinsic molecular properties, particularly the internal bond distances, with a smaller contribution from the hydrogen bond network. This suggest that, for its spectral properties, liquid water behaves as a collection of mostly isolated molecules, with a minor influence from its intermolecular structure. Moreover, the methodology employed in this work can be extended to the electronic properties of solvated ions, which are being studied with GW methodologies and also with implicit solvents\cite{ISI:000497260300041, ISI:000377643300002, Phame1603210, PhysRevLett.111.087801}.

\begin{acknowledgement}
The authors would like to thank the financial support of CNPq, PRH/ANP, the Swiss National Science Foundation (SNSF) through its National Centre of Competence in Research (NCCR) MARVEL, grant 200021-179138, and the National Laboratory for Scientific Computing (LNCC/MCTI,
Brazil) for providing HPC resources on the SDumont supercomputer.
\end{acknowledgement}

\begin{suppinfo}
This manuscript contains a supporting info file, which contains: Figure S1, with aligned edge levels; Table S1, which the edge level changes when going from classical to quantum nuclei; Figure S2, with the band gap variation as a function of a rescaled O-H bond length for one snapshot; Figure S3, with the band edge values as a function of a rescaled O-H bond length for one snapshot; Figure S4, with the band gap variations for different rescaled properties of a water dimer.
\end{suppinfo}

\begin{tocentry}
\includegraphics[width=0.99\textwidth]{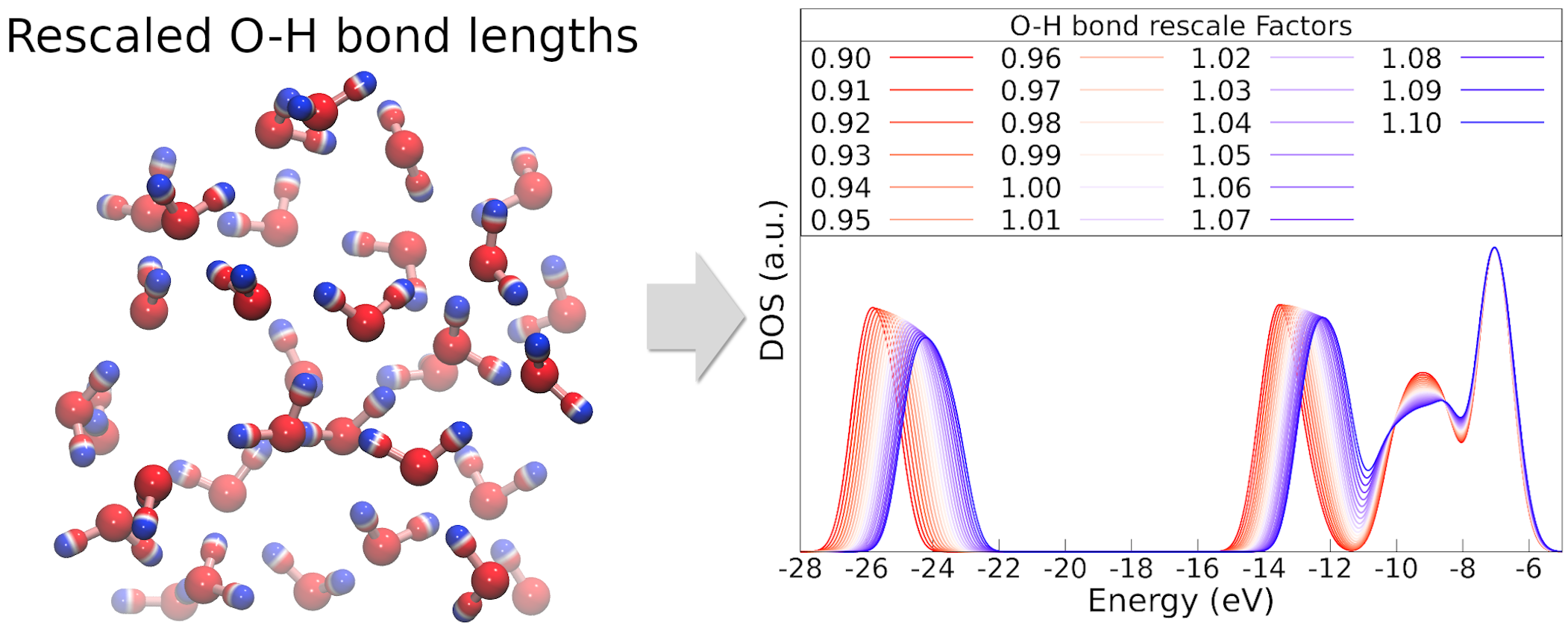}
\end{tocentry}

\bibliography{water-kp}

\end{document}